\documentclass[conference]{IEEEtran}
\IEEEoverridecommandlockouts

\usepackage{cite}
\usepackage{amsmath,amssymb,amsfonts}
\usepackage{algorithmic}
\usepackage{graphicx}
\usepackage{textcomp}
\usepackage{multirow}
\usepackage{tabularx}
\usepackage{booktabs}
\usepackage[table]{xcolor}
\usepackage{url}
\def\BibTeX{{\rm B\kern-.05em{\sc i\kern-.025em b}\kern-.08em
    T\kern-.1667em\lower.7ex\hbox{E}\kern-.125emX}}
\begin{document}

\title{BadViM: Backdoor Attack against Vision Mamba}

\author{\IEEEauthorblockN{Yinghao Wu}
\IEEEauthorblockA{\textit{College of Computer Science and Technology} \\
\textit{Nanjing University of Aeronautics and Astronautics}\\
Nanjing, China \\
wyh@nuaa.edu.cn}
\and
\IEEEauthorblockN{Liyan Zhang}
\IEEEauthorblockA{\textit{College of Computer Science and Technology} \\
\textit{Nanjing University of Aeronautics and Astronautics}\\
Nanjing, China \\
zhangliyan@nuaa.edu.cn}
}

\newcommand{\mS}{\mathbf{S}}
\newcommand{\mK}{\mathbf{K}}
\newcommand{\mV}{\mathbf{V}}
\newcommand{\mZ}{\mathbf{Z}}
\newcommand{\vy}{\mathbf{y}}
\newcommand{\mQ}{\mathbf{Q}}
\newcommand{\vh}{\mathbf{h}}
\newcommand{\mA}{\mathbf{A}}
\newcommand{\mB}{\mathbf{B}}
\newcommand{\vx}{\mathbf{x}}
\newcommand{\mC}{\mathbf{C}}
\newcommand{\mD}{\mathbf{D}}
\maketitle

\begin{abstract}
Vision State Space Models (SSMs), particularly architectures like Vision Mamba (ViM), have emerged as promising alternatives to Vision Transformers (ViTs).
However, the security implications of this novel architecture, especially their vulnerability to backdoor attacks, remain critically underexplored.
Backdoor attacks aim to embed hidden triggers into victim models, causing the model to misclassify inputs containing these triggers while maintaining normal behavior on clean inputs.
This paper investigates the susceptibility of ViM to backdoor attacks by introducing BadViM, a novel backdoor attack framework specifically designed for Vision Mamba.
The proposed BadViM leverages a Resonant Frequency Trigger (RFT) that exploits the frequency sensitivity patterns of the victim model to create stealthy, distributed triggers.
To maximize attack efficacy, we propose a Hidden State Alignment loss that strategically manipulates the internal representations of model by aligning the hidden states of backdoor images with those of target classes.
Extensive experimental results demonstrate that BadViM achieves superior attack success rates while maintaining clean data accuracy.
Meanwhile, BadViM exhibits remarkable resilience against common defensive measures, including PatchDrop, PatchShuffle and JPEG compression, which typically neutralize normal backdoor attacks.
\end{abstract}


\section{Introduction}\label{sec:intro}
Vision State-Space Models (SSMs), exemplified by Vision Mamba (ViM \cite{ViM_ZhuL0W0W24} and VMamba \cite{VMamba_LiuTZYX0YJ024}), have emerged as promising alternatives to ViTs by achieving linear computational complexity.
ViM processes an image by first converting it into a sequence of patches, similar to ViT, but then analyzes this sequence using a serial state accumulation process. 
Information from previous patches is compressed into a fixed-size hidden state vector, which is recurrently updated as the model scans through the sequence.
This transition from parallel attention to serial state processing is not merely a performance optimization.
This fundamental restructuring of how a model perceives and reasons about visual information also creates a distinct attack surface, demanding a re-evaluation of established threats like backdoor attacks \cite{badnet_abs-1708-06733, wu2025cuba} that have already proven effective against prior architectures such as ViTs \cite{YouAreCatching2023Yuan, subramanya2022backdoor, subramanya2024closer, zheng2023trojvit}.

Recent research \cite{QRDBA_NagaonkarTM25} indicate that traditional backdoor attacks \cite{badnet_abs-1708-06733, WanetImperceptibleWarpingBased_Nguyen2021, ReflectionBackdoor_LiuM0020, wu2025cuba} show limited effectiveness \cite{QRDBA_NagaonkarTM25} against ViM due to the ``memory attenuation'' phenomenon, where the sequential state-update mechanism progressively attenuates spatially isolated perturbations.
While ViM may gain resilience against certain types of attacks due to its unique processing style, it simultaneously introduces a new, potentially more severe, centralized vulnerability. The distributed and complex flow of interactions in a ViT's attention mechanism presents a diffuse target for an attacker. In contrast, the ViM architecture compresses all historical information into a single, compact hidden state vector.
While this information bottleneck is instrumental for the efficiency of model and may help in forgetting or washing out isolated, noisy signals like a single malicious patch, it also implies a critical concentration of risk.
If an attacker can devise a method to reliably control this single vector, they effectively seize control of the model's entire perceptual understanding of the input sequence.
The attack surface has thus migrated from a distributed system (attention weights) to a centralized one (hidden state), and such centralized systems are notoriously more fragile if a critical vulnerability can be identified and exploited.

In this paper, we propose BadViM, a novel backdoor attack framework specifically designed for Vision Mamba architectures.
BadViM exploits the centralized hidden state vulnerability of ViM through \textbf{Resonant Frequency Trigger (RFT)} and \textbf{Hidden State Alignment}.
Specifically, we identify model-specific \textit{resonant frequencies} and inject subtle perturbations exclusively within these spectral regions, ensuring sustained stimulation throughout sequence processing.
The Hidden State Alignment loss function directly manipulates the model's internal representations by aligning poisoned inputs' hidden states with target class manifolds.
The main contributions are summarized as follows:
\begin{itemize}
\item We introduce BadViM, the first backdoor attack framework combining frequency-domain triggers with hidden state manipulation for Vision State Space Models.
We demonstrate superior attack performance with enhanced stealth and robustness than state-of-the-art attacks.
\item We provide a theoretical analysis, interpreting ViM as a linear attention model, to reveal how its \textit{forget gate} architecture inherently resists localized patch attacks while remaining vulnerable to our persistent, distributed trigger.
\end{itemize}

\begin{figure*}[htbp]
    \centering
    \includegraphics[width=1.0\linewidth]{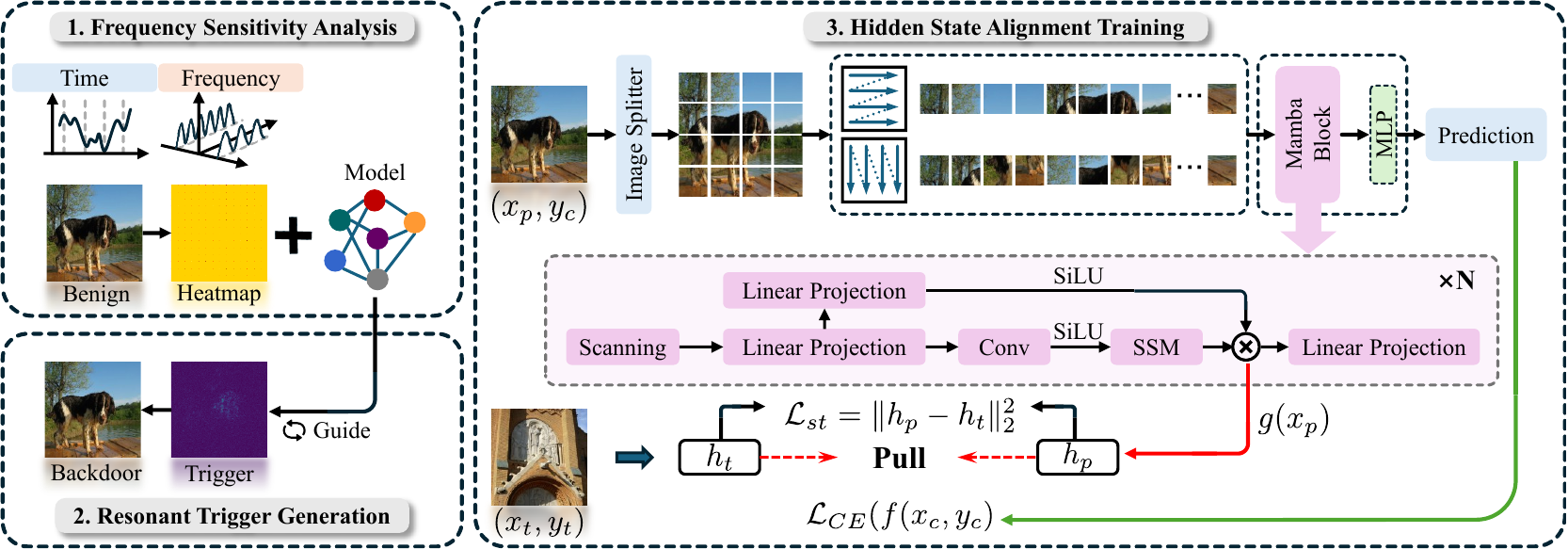}
    \caption{The illustration of the proposed BadViM attack framework.}
    \label{fig:main}
\end{figure*}

\section{Threat Model}\label{sec:threat_model}
Following existing backdoor attacks \cite{WanetImperceptibleWarpingBased_Nguyen2021, DeepFeatureSpace_Cheng2021, BlindBackdoorsDeep_Bagdasaryan2021, LIRA_Doan2021, BppAttackStealthyEfficient_Wang2022} We consider a training-manipulation backdoor attack where the adversary has comprehensive control over the model training process.

\textbf{Attacker's Goal}: The attacker aims to embed a persistent backdoor into a Vision Mamba model that satisfies three criteria: 1) \textit{Effectiveness}: triggered inputs are misclassified to a target label with high Attack Success Rate (ASR); 2) \textit{Stealthiness}: triggers remain visually imperceptible and model performance on clean data is preserved (high Clean Data Accuracy); 3) \textit{Robustness}: the backdoor persists under common defenses and data transformations.

\textbf{Attacker's Capabilities}: The attacker controls the entire training pipeline, including loss function design, training data modification, and access to intermediate model representations.
This enables manipulation of both input-output relationships and internal hidden state dynamics.

\section{The proposed BadViM framework}\label{sec:badvim}

\subsection{Preliminaries}
Vision Mamba processes input images through selective State-Space Model (SSM) dynamics. An input image is partitioned into $N$ patches and projected into token embeddings $\mathbf{x} = (x_1, \ldots, x_N)$. The core SSM operation for the $i$-th token is:
\begin{equation}\label{eq:ssm_discrete}
h(i) = \overline{\mathbf{A}}_i h(i-1) + \overline{\mathbf{B}}_i x_i
\end{equation}
where $h(i)$ is the hidden state and parameters $\overline{\mathbf{A}}_i, \overline{\mathbf{B}}_i$ are input-dependent for selective processing. After processing all tokens, the final hidden state $h(N)$ serves as the global representation for classification.

\subsection{Overview}
BadViM operates through an iterative training process that alternates between trigger generation and model training. Given a pre-trained or partially trained Vision Mamba model, the framework first analyzes the model's frequency sensitivity to identify optimal perturbation regions, then generates frequency-domain triggers targeting these sensitive bands. During retraining or fine-tuning (backdoor injection), a composite loss function enforces both correct target label prediction and internal hidden state alignment, ensuring robust backdoor embedding.

The attack proceeds in iterations: 1) estimate frequency sensitivity heatmap for the current model, 2) generate triggers based on identified resonant frequencies, 3) create poisoned samples and update the model using the composite loss, 4) repeat until convergence. This co-evolution of triggers and model weights maximizes attack effectiveness while maintaining stealth.

\subsection{Resonant Frequency Trigger Generation}
\textbf{Frequency Sensitivity Analysis.} Vision Mamba models exhibit varying sensitivity to different frequency components due to their architectural inductive biases and training data characteristics.
Inspired by FreqBack \cite{FreqBack_0001ZWLY25}, we estimate this sensitivity through a frequency heatmap $S(u,v)$ that quantifies how model predictions change when specific frequency components are perturbed.

For each frequency component $(u,v)$ in the 2D Fourier domain, we create a pure sinusoidal perturbation $P_{u,v}$ and apply it to clean validation images: $I'_{u,v} = I + \epsilon \cdot P_{u,v}$. The sensitivity is measured as:
\begin{equation}\label{eq:frequency_sensitivity}
S(u,v) = \frac{1}{|\mathcal{V}|} \sum_{I \in \mathcal{V}} [\ell(f(I'_{u,v}), y) - \ell(f(I), y)]
\end{equation}
where $\mathcal{V}$ is the validation set, $\ell$ is the cross-entropy loss, and $\epsilon$ controls perturbation magnitude.

\textbf{Trigger Construction.} The trigger generation targets the most sensitive frequency regions identified by the heatmap. We select the top-$k\%$ frequencies with highest sensitivity scores to form a binary mask $M_{\text{freq}}$. A complex-valued noise pattern $N_{\text{freq}}$ is generated from a zero-mean Gaussian distribution and masked to create the frequency-domain trigger:
\begin{equation}\label{eq:frequency_trigger}
\delta_{\text{freq}} = N_{\text{freq}} \odot M_{\text{freq}}
\end{equation}
The spatial-domain trigger is obtained via inverse Fourier transform: $\delta = \mathcal{F}^{-1}(\delta_{\text{freq}})$. Poisoned images are generated as: $x_p = \text{clip}(x_c + \delta, 0, 1)$.

\subsection{Hidden State Alignment Training}
A core component of BadViM is its training objective, which directly manipulates the model's internal feature manifold by simultaneously enforcing correct behavior on clean data and malicious behavior on poisoned data.

\textbf{Target State Definition.} Let $g(\cdot)$ be the Vision Mamba backbone that maps an input image $x$ to its final global hidden state vector, $h_N = g(x)$. For a target class $y_t$, we define its representative hidden state vector $h_t$ as the centroid of the final hidden states of all clean samples belonging to that class:
\begin{equation}\label{eq:target_state}
h_t = \mathbb{E}_{x_c \sim \mathcal{D}_t} [g(x_t)]
\end{equation}
where $\mathcal{D}_t$ is the data distribution for the clean target class. In practice, this centroid is estimated and dynamically updated using mini-batches during training.

\textbf{Composite Loss Function.} The backdoor injection process is driven by a composite loss function separating the learning objectives for clean and poisoned samples. For a clean sample $(x_c, y_c)$, we generate its poisoned counterpart $x_p = x_c + \delta$. The total loss for the model update is $\mathcal{L}_{tot} = \mathcal{L}_{c} + \mathcal{L}_{p}$, where the clean loss $\mathcal{L}_{c} = \mathcal{L}_{CE}(f(x_c), y_c)$ ensures performance on benign data is not degraded.

The poison loss $\mathcal{L}_{p}$ is applied to the triggered sample $x_p$ and is responsible for implanting the backdoor. It forces both misclassification to the target label $y_t$ and alignment of the internal representation with the target class manifold:
\begin{equation}\label{eq:poison_loss}
\mathcal{L}_{p} = \mathcal{L}_{CE}(f(x_c), y_c) + \lambda \cdot \mathcal{L}_{st}(g(x_p), h_t)
\end{equation}
where $f(\cdot)$ is the full model, $\lambda$ is a hyperparameter, and the state alignment loss $\mathcal{L}_{st}$ is the squared Euclidean distance between the hidden states, i.e., $\mathcal{L}_{st} = \| g(x_p) - h_t \|_2^2$.

By optimizing this composite loss, BadViM learns a dual behavior. Driven by $\mathcal{L}_{c}$, it processes benign inputs correctly. However, when the RFT is detected, the model is guided by $\mathcal{L}_{p}$ to produce the target output while simultaneously steering its final hidden state $h_N$ into the geometric heart of the target class's feature manifold.

\section{Theoretical Analysis} \label{sec:theory}
Recent work \cite{demystirfymambainvision_HanWXHPGSSZ024} reveals that Vision Mamba's Selective State Space Model can be interpreted as a specialized form of recurrent linear attention, providing insights into its vulnerabilities.

\subsection{Vision Mamba as Recurrent Linear Attention}
The recurrent state update of single-head causal linear attention is:
\begin{equation} \label{eq:linear_attention_recurrent}
\begin{split}
    \mS(i) &= \mS(i-1) + \mK(i)^{\top}\mV(i) \\
    \mZ(i) &= \mZ(i-1) + \mK(i)^{\top} \\
    \vy(i) &= \frac{\mQ(i)\mS(i)}{\mQ(i)\mZ(i)}
\end{split}
\end{equation}
where $\mQ(i), \mK(i), \mV(i)$ are the query, key, and value for the $i$-th token.

Mamba's S6 block (Selective SSM) operates as:
\begin{equation} \label{eq:mamba_recurrent}
\begin{split}
    \vh(i) &= \widetilde{\mA}(i) \odot \vh(i-1) + \mB(i) (\mathbf{\Delta}(i) \odot \vx(i)) \\
    \vy(i) &= \mC(i) \vh(i) + \mD \odot \vx(i)
\end{split}
\end{equation}

Comparing these formulations reveals key distinctions:
\begin{itemize}
    \item \textbf{Forget Gate ($\widetilde{\mA}(i)$):} Controls information decay from previous states.
    \item \textbf{Input Gate ($\mathbf{\Delta}(i)$):} Modulates current input contribution.
    \item \textbf{No Normalization:} Unlike linear attention, making state updates more sensitive to input magnitude.
\end{itemize}

\subsection{Architectural Vulnerabilities of ViM}
The BadViM framework is engineered to systematically exploit the core components of this linear attention-like structure.

\subsubsection{The Resonant Frequency Trigger (RFT) as a Persistent State Corruptor}
The \textit{memory forgetting} hypothesis, which suggests ViM can ``wash out'' localized triggers, is a direct consequence of the forget gate $\widetilde{\mA}(i)$.
A single-patch trigger creates a large, transient change in the state update, but the forget gate quickly attenuates this signal in subsequent steps.

Resonant Frequency Trigger (RFT) is designed to counteract this. By distributing a low-magnitude trigger across the \textit{entire} image, we ensure that a malicious signal is present in \textit{every} input token $\vx(i)$.
Let the poisoned input be $\vx'(i) = \vx(i) + \delta(i)$, where $\delta(i)$ is the portion of the spatial trigger in patch $i$. The state update becomes:
\begin{equation}\label{eq:poisoned_state_detailed}
\vh(i) = \widetilde{\mA}(i) \odot \vh(i-1) + \mB(i) (\mathbf{\Delta}(i) \odot (\vx(i) + \delta(i)))
\end{equation}
The RFT provides a persistent, low-magnitude perturbation term, $\mB(i) (\mathbf{\Delta}(i) \odot \delta(i))$, at every step of the recurrence.
The memory forgetting mechanism fails because there is nothing to forget; the malicious signal is continuously re-injected.
The model's sensitivity to specific \textit{resonant} frequencies means that even a small $\delta(i)$ can have a disproportionately large impact on the state update, allowing the backdoor to accumulate influence stealthily without being attenuated.

\subsubsection{Hidden State Alignment: Direct Control of the Accumulated State}
Traditional backdoor attacks only supervise the final output label. BadViM's Hidden State Alignment goes deeper by directly manipulating the final accumulated state vector $h(N)$.
From the linear attention perspective, the final state $h(N) = g(x)$ is the result of the entire recurrent accumulation process:
\begin{equation}\label{eq:state_accumulation}
h(N) = \sum_{i=1}^{N} \left( \left( \prod_{j=i+1}^{N} \widetilde{\mA}(j) \right) \odot \mB(i) (\mathbf{\Delta}(i) \odot \vx(i)) \right)
\end{equation}
\noindent which is a simplified representation of the associative scan.
The proposed state alignment loss $\mathcal{L}_{st} = \| g(x_p) - h_t \|_2^2$, forces the entire accumulation process to converge to a specific point in the feature space $h_t$. Instead of just hoping the trigger pushes the final output over a decision boundary, we are explicitly teaching the model a malicious state-space trajectory.

\begin{table}[t]
  \centering
  \caption{Performance comparison of different attack methods across various vision backbones. CDA denotes Clean Data Accuracy (\%), and ASR denotes Attack Success Rate (\%).}
  \setlength{\tabcolsep}{4pt} 
  \begin{tabular}{llcccc}
    \toprule
    \multirow{2}[2]{*}{Model} & \multirow{2}[2]{*}{Attack} & \multicolumn{2}{c}{CIFAR-10} & \multicolumn{2}{c}{ImageNet-1k (subset)} \\
    \cmidrule(lr){3-4} \cmidrule(lr){5-6}
     & & CDA & ASR & CDA & ASR \\
    \midrule
    \multirow{4}{*}{ResNet-18} & None & 95.13 & - & 78.51 & - \\
     & Lee \textit{et al.} \cite{abs-2408-11679} & - & - & 69.67 & 99.24 \\
     & QRDBA \cite{QRDBA_NagaonkarTM25} & 83.11 & 55.93 & 94.00 & 68.80 \\
     & {\cellcolor [HTML]{EFEFEF}BadViM (Ours)} & {\cellcolor [HTML]{EFEFEF}94.97} & {\cellcolor [HTML]{EFEFEF}\textbf{99.98}} & {\cellcolor [HTML]{EFEFEF}78.21} & {\cellcolor [HTML]{EFEFEF}\textbf{99.75}} \\
    \midrule
    \multirow{3}{*}{ViM-T} & None & 87.95 & - & 78.65 & - \\
     & Lee \textit{et al.} \cite{abs-2408-11679} & - & - & 77.11 & 97.89 \\
     & {\cellcolor [HTML]{EFEFEF}BadViM (Ours)} & {\cellcolor [HTML]{EFEFEF}88.01} & {\cellcolor [HTML]{EFEFEF}\textbf{99.79}} & {\cellcolor [HTML]{EFEFEF}78.62} & {\cellcolor [HTML]{EFEFEF}\textbf{99.94}} \\
    \midrule
    \multirow{4}{*}{VManba-T} & None & 86.73 & - & 81.76 & - \\
     & BadScan \cite{BadScan_abs-2411-17283} & 94.00 & 8.00 & 93.00 & 6.00 \\
     & QRDBA \cite{QRDBA_NagaonkarTM25} & 88.00 & 44.43 & 75.00 & 40.00 \\
     & {\cellcolor [HTML]{EFEFEF}BadViM (Ours)} & {\cellcolor [HTML]{EFEFEF}86.12} & {\cellcolor [HTML]{EFEFEF}\textbf{99.54}} & {\cellcolor [HTML]{EFEFEF}81.24} & {\cellcolor [HTML]{EFEFEF}\textbf{99.97}} \\
    \midrule
    \multirow{3}{*}{MambaOut-T} & None & 88.14 & - & 83.49 & - \\
     & Lee \textit{et al.} \cite{abs-2408-11679} & 86.37 & 97.86 & 78.68 & 98.32 \\
     & {\cellcolor [HTML]{EFEFEF}BadViM (Ours)} & {\cellcolor [HTML]{EFEFEF}87.46} & {\cellcolor [HTML]{EFEFEF}\textbf{98.03}} & {\cellcolor [HTML]{EFEFEF}82.57} & {\cellcolor [HTML]{EFEFEF}\textbf{98.89}} \\
    \bottomrule
  \end{tabular}
  \label{tab:main_results}
\end{table}

\section{Experiments}\label{sec:exp}
\subsection{Experimental Settings}
\textbf{Datasets and Models.} We evaluate BadViM on CIFAR-10 and ImageNet-1K (100-class subset) using ViM-Tiny \cite{ViM_ZhuL0W0W24}, VMamba-Tiny \cite{VMamba_LiuTZYX0YJ024} and MambaOut-Tiny \cite{mambaout_yu2025mambaout} for reference.
MambaOut removes the SSM module and therefore works more like a Gated CNN model.
To evaluate the proposed attack against ViM variants (e.g., MambaOut), we also conduct the experiments on MambaOut-Tiny \cite{mambaout_yu2025mambaout}.
All weights of pre-trained model are downloaded from the official Github repositories (ViM-Tiny\footnote{\url{https://github.com/hustvl/Vim}}, VMamba-Tiny\footnote{\url{https://github.com/MzeroMiko/VMamba}} and MambaOut-Tiny\footnote{\url{https://github.com/yuweihao/MambaOut}}).
\textbf{Evaluation Metrics.}
Metrics include Attack Success Rate (ASR) and Clean Data Accuracy (CDA), which indicates the classification accuracy of the backdoored model on backdoor images and clean images respectively.

\begin{figure}[htbp]
    \centering
    \includegraphics[width=1.0\linewidth]{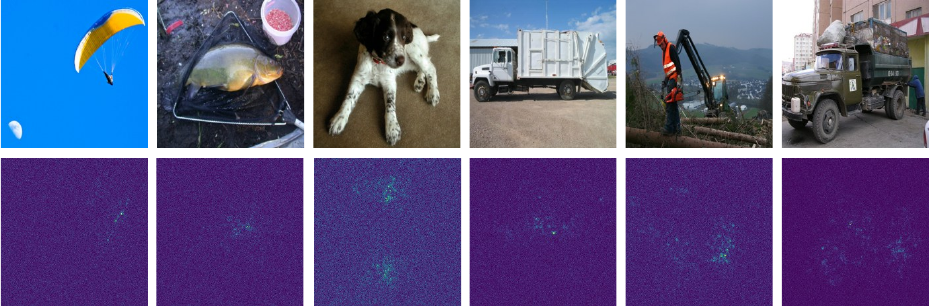}
    \caption{Examples of clean images and corresponding frequency triggers on ImageNet-1k. First row and second row are clean images and corresponding triggers respectively.}
    \label{fig:trigger}
\end{figure}

\subsection{Main Results}
The experimental results are reported in Table \ref{tab:main_results}.
The examples of generated triggers and clean images are shown in Fig. \ref{fig:trigger}.
It shows that our method achieves the highest ASR on all models and datasets while maintaining a negligible loss in CDA.
Specifically, BadViM achieves superior performance with near-perfect ASR (higher than 99.8\%) across all datasets while maintaining minimal CDA degradation.
We also compare against three existing attacks for ViM: Lee \textit{et al.} \cite{abs-2408-11679}, BadScan \cite{BadScan_abs-2411-17283} and QRDBA \cite{QRDBA_NagaonkarTM25} by using a 10\% poisoning rate.
It demonstrates that the proposed BadViM can achieves higher attack success rate than all three existing attacks.
Note that, BadScan \cite{BadScan_abs-2411-17283} solely relies on the architecture defect of ViM, and hence achieves a relatively lower ASR.

\subsection{Robustness against Defenses}
We evaluate BadViM against several common defenses.
As shown in Table \ref{tab:defense}, our attack is significantly more robust than the baseline.
For defenses like PatchDrop and PatchShuffle, which target localized triggers, BadViM's Attack Success Rate (ASR) remains above 99\%.
The reason for this resilience is twofold. First, the distributed Resonant Frequency Trigger (RFT) ensures a malicious signal is present in most patches. Second, and more importantly, due to the Hidden State Alignment training, the model learns to steer its hidden state vector $h(i)$ towards the target manifold even with incomplete trigger signals.
When some patches are dropped or shuffled, the cumulative input from the remaining RFT-corrupted patches is still potent enough to guide the state accumulation process to the learned malicious endpoint, making the defense ineffective.
Similarly, under JPEG compression, BadViM maintains a high ASR of 98.54\%. This is because the RFT is not embedded in arbitrary frequency bands, but specifically in the model's \textit{resonant frequencies}, where the bands it is most sensitive to for classification. These critical frequencies are often in the low-to-mid range and are naturally preserved during JPEG's high-frequency quantization process. The trigger therefore survives compression and remains effective.
In summary, BadViM effectively bypasses these common defenses, demonstrating its practical robustness.

\begin{table}[t]
  \centering
  \caption{Attack Success Rate (\%) of BadViM compared with the baseline from Lee \textit{et al.} \cite{abs-2408-11679} against various defenses on the ImageNet-1k subset.}
  \begin{tabular}{lcccc}
    \toprule
    Attack Method & No Defense & PatchDrop & PatchShuffle & JPEG \\
    \midrule
    Lee \textit{et al.} \cite{abs-2408-11679} & 97.89 & 94.06 & 93.14 & 95.98 \\
    \textbf{BadViM (Ours)} & \textbf{99.98} & \textbf{99.12} & \textbf{99.85} & \textbf{98.54} \\
    \bottomrule
  \end{tabular}
  \label{tab:defense}
\end{table}

\section{Conclusion}
This paper proposes BadViM, the first backdoor attack framework specifically designed for Vision Mamba architectures. By exploiting the centralized hidden state vulnerability through resonant frequency triggers and hidden state alignment, BadViM achieves superior attack performance while maintaining stealth. Our theoretical analysis reveals the linear attention analogy of ViM's vulnerability, providing insights for both attack and defense development. The exceptional robustness against state-of-the-art defenses demonstrates the practical threat posed by BadViM, highlighting the need for new security measures in Vision State-Space Models.

\bibliographystyle{IEEEtran}
\bibliography{ref}

\end{document}